\documentclass{article}
\usepackage{amssymb}
\usepackage{amsfonts}
\usepackage{amsmath}

\input{tcilatex}

\begin{document}

\begin{center}
\bigskip

{\large THE GRAVITATIONAL CONTENT OF LORENTZIAN COMPLEX STRUCTURES}

{\large \ }\bigskip

C. N. Ragiadakos

Pedagogical Institute

Mesogion 396, Agia Paraskevi, TK 15341, Greece

email: crag@pi-schools.gr

\bigskip

\textbf{ABSTRACT}

The definition of a positive energy is investigated in a renormalizable
4-dimensional generally covariant model, which depends on the lorentzian
complex structure and not the metric of spacetime. The gravitational content
of the lorentzian complex structures is revealed by identifying the
spacetime with special 4-dimensional surfaces of the $G_{2,2}$ Grassmannian
manifold. The lorentzian complex structure is found to be a codimension-4 CR
structure and its classification is studied using the Chern-Moser and Cartan
methods. The spacetime metric is found to be a Fefferman-like metric of this
codimension-4 CR structure. The open CR manifolds "hanging" from the points
of the $U(2)$ characteristic boundary of the $SU(2,2)$ classical domain
belong into representations of the Poincar\'{e} group and are related to the
particle spectrum of the model.
\end{center}

\pagebreak

\bigskip

\bigskip

\bigskip

{\LARGE Contents}

\textbf{1. INTRODUCTION}

\qquad

\textbf{2. AN\ ENERGY\ DEFINITION\ FOR\ THE\ MODEL}

\qquad

\textbf{3. THE }$G_{2,2}$\textbf{\ DESCRIPTION OF THE LORENTZIAN COMPLEX
STRUCTURE}

\qquad

\textbf{4. CLASSIFICATION OF LORENTZIAN COMPLEX STRUCTURES}

\qquad 4.1 The Chern-Moser normal form method

\qquad 4.2 The Cartan connection method

\qquad

\textbf{5. A FEFFERMAN-LIKE METRIC}

\bigskip

\textbf{6. ON U(2) AND POINCARE SYMMETRIES}

\pagebreak

\renewcommand{\theequation}{\arabic{section}.\arabic{equation}}

\section{INTRODUCTION}

\setcounter{equation}{0}

After the recent failure of ATLAS and CMS experiments to find minimal
supersymmetry effects and (large) higher spacetime dimensions, doubts on the
physical relevance of the superstring model start to appear. Despite these
experimental difficulties, the proponents of the string model do not give
up, because they think that it is the unique quantum mechanically
self-consistent model, which includes gravity. In fact, it is not unique!
Long time ago\cite{RAG1988},\cite{RAG1990},\cite{RAG1991},\cite{RAG1992} I
wrote down and I quantized the following simple 4-dimensional
Yang-Mills-like action, which depends on the lorentzian complex structure
and not the metric of the spacetime. 
\begin{equation}
\begin{array}{l}
I_{G}=\int d^{4}\!z\sqrt{-g}g^{\alpha \widetilde{\alpha }}g^{\beta 
\widetilde{\beta }}\ F_{\!j\alpha \beta }F_{\!j\widetilde{\alpha }\widetilde{%
\beta }}+c.\ c.=2\int d^{4}\!z\ F_{\!j01}F_{\!j\widetilde{0}\widetilde{1}%
}+c.\ c. \\ 
\\ 
F_{jab}=\partial _{a}A_{jb}-\partial _{a}A_{jb}-\gamma \,f_{jik}A_{ia}A_{kb}%
\end{array}
\label{i1}
\end{equation}

Lorentzian complex structures have been introduced\cite{FLAHE1974},\cite%
{FLAHE1976} by Flaherty in order to study spacetimes with two geodetic and
shear free congruences. Using the ordinary null tetrad $(\ell _{\mu
},\,n_{\mu },\,m_{\mu },\,\overline{m}_{\mu })$, the lorentzian metric $%
g_{\mu \nu }$ and the complex structure $J_{\mu }^{\;\nu }$ take the form

\begin{equation}
\begin{array}{l}
g_{\mu \nu }=\ell _{\mu }n_{\nu }+n_{\mu }\ell _{\nu }-m_{{}\mu }\overline{m}%
_{\nu }-\overline{m}_{\mu }m_{\nu } \\ 
\\ 
J_{\mu }^{\;\nu }=i(\ell _{\mu }n^{\nu }-n_{\mu }\ell ^{\nu }-m_{\mu }%
\overline{m}^{\nu }+\overline{m}_{\mu }m^{\nu })%
\end{array}
\label{i2}
\end{equation}%
Notice that the lorentzian complex structure $J_{\mu }^{\;\nu }$ is a
complex tensor, unlike the ordinary euclidean complex structure, which is a
real tensor. This complex structure is integrable if

\begin{equation}
\begin{array}{l}
(\ell ^{\mu }m^{\nu }-\ell ^{\nu }m^{\mu })(\partial _{\mu }\ell _{\nu
})=0\;\;\;\;,\;\;\;\;(\ell ^{\mu }m^{\nu }-\ell ^{\nu }m^{\mu })(\partial
_{\mu }m_{\nu })=0 \\ 
\\ 
(n^{\mu }m^{\nu }-n^{\nu }m^{\mu })(\partial _{\mu }n_{\nu
})=0\;\;\;\;,\;\;\;\;(n^{\mu }m^{\nu }-n^{\nu }m^{\mu })(\partial _{\mu
}m_{\nu })=0%
\end{array}
\label{i3}
\end{equation}%
That is when the spin coefficients $\kappa ,\ \sigma ,\ \lambda ,\ \nu $
vanish, which implies that the real vectors $\ell ^{\mu }$\ and $n^{\mu }$
define geodetic and shear free congruences. Then Frobenius theorem states
that there are four independent complex functions $(z^{\alpha },\;z^{%
\widetilde{\alpha }})$,\ $\alpha =0,\ 1$ , such that

\begin{equation}
\begin{array}{l}
dz^{\alpha }=f_{\alpha }\ \ell _{\mu }dx^{\mu }+h_{\alpha }\ m_{\mu }dx^{\mu
}\;\;\;\;,\;\;\;dz^{\widetilde{\alpha }}=f_{\widetilde{\alpha }}\ n_{\mu
}dx^{\mu }+h_{\widetilde{\alpha }}\ \overline{m}_{\mu }dx^{\mu } \\ 
\\ 
\ell =\ell _{\alpha }dz^{\alpha }\;\;\;\;,\;\;\;m=m_{\alpha }dz^{\alpha } \\ 
\\ 
n=n_{\widetilde{\alpha }}dz^{\widetilde{\alpha }}\;\;\;\;,\;\;\;m=m_{%
\widetilde{\alpha }}dz^{\widetilde{\alpha }} \\ 
\end{array}
\label{i4}
\end{equation}

These four functions are the structure coordinates of the (integrable)
complex structure. In the present case of lorentzian spacetimes the
coordinates $z^{\widetilde{\alpha }}$ are not complex conjugate of $%
z^{\alpha }$, because $J_{\mu }^{\;\nu }$ is no longer a real tensor, like
the ordinary complex structures. We always have $\overline{z^{a}}%
=f^{a}(z^{b})$, while the condition $dz^{\widetilde{0}}\wedge dz^{\widetilde{%
0}}\wedge dz^{0}\wedge dz^{1}\neq 0$ has to be imposed.

Using the null tetrad, the generally covariant form of model action takes
the form 
\begin{equation}
\begin{array}{l}
I_{G}=\int d^{4}\!x\ \sqrt{-g}\ \left\{ \left( \ell ^{\mu }m^{\rho
}F_{\!j\mu \rho }\right) \left( n^{\nu }\overline{m}^{\sigma }F_{\!j\nu
\sigma }\right) +\left( \ell ^{\mu }\overline{m}^{\rho }F_{\!j\mu \rho
}\right) \left( n^{\nu }m^{\sigma }F_{\!j\nu \sigma }\right) \right\} \\ 
\\ 
F_{j\mu \nu }=\partial _{\mu }A_{j\nu }-\partial _{\nu }A_{j\mu }-\gamma
\,f_{jik}A_{i\mu }A_{k\nu }%
\end{array}
\label{i5}
\end{equation}%
with the following term of Lagrange multipliers 
\begin{equation}
\begin{array}{l}
I_{C}=\int d^{4}\!x\ \sqrt{-g}\{\phi _{0}(\ell ^{\mu }m^{\nu }-\ell ^{\nu
}m^{\mu })(\partial _{\mu }\ell _{\nu })+\phi _{1}(\ell ^{\mu }m^{\nu }-\ell
^{\nu }m^{\mu })(\partial _{\mu }m_{\nu })+ \\ 
\\ 
\qquad +\phi _{\widetilde{0}}(n^{\mu }\overline{m}^{\nu }-n^{\nu }\overline{m%
}^{\mu })(\partial _{\mu }n_{\nu })+\phi _{\widetilde{1}}(n^{\mu }\overline{m%
}^{\nu }-n^{\nu }\overline{m}^{\mu })(\partial _{\mu }\overline{m}_{\nu
})+c.conj.\}%
\end{array}
\label{i6}
\end{equation}%
which impose the integrability conditions of the lorentzian complex
structure. These terms are essential, because they assure the metric
independence of the action leading to its renormalizability\cite{RAG2008a}.
In brief, this model is a conventional lagrangian generally covariant model
which is renormalizable because of its increased symmetry.

The physical content of the model has been studied in my previous works\cite%
{RAG1999},\cite{RAG2008b},\cite{RAG2010}. In this work we focus on the
mathematical methods which could be used for the definition of a positive
energy and the related classification of the lorentzian complex structures.
In section II, I specify the general class of metrics, which are compatible
with a lorentzian complex structure. I have not yet found an appropriate
definition of the energy quantity of the model. A formal definition of the
energy may be undertaken using the existence of a coordinate system, where
the ordinary contracted derivative of the Einstein tensor vanishes. But this
coordinate system has to be related to the geodetic coordinates of $\ell
^{\mu }$ and $n^{\mu }$ which transform according to a Poincar\'{e} group.
This defined energy, suggested by the great success of the Einstein general
relativity, must be proved to be positive.

The Poincar\'{e} group naturally emerges if the spacetime is described as a
surface of $G_{2,2}$ in section III and a codimension-4 CR manifold in
section IV, where the classification of lorentzian complex structures is
undertaken. In section V, I find that the spacetime class of metrics
compatible with a lorentzian complex structure are Fefferman-like metrics of
the corresponding codimension-4 CR structure. This direct relation of the
"particles" of the model with the formidable mathematical machinery of
bounded domains, their CR-boundaries and the natural emergence of the Poincar%
\'{e} group may permit us to define the energy quantity.

\section{AN ENERGY DEFINITION FOR THE MODEL}

\setcounter{equation}{0}

The integrable complex structure is determined by a null tetrad up to the
following transformations%
\begin{equation}
\begin{tabular}{l}
$\ell _{\mu }^{\prime }=\Lambda \ell _{\mu }\quad ,\quad \ell ^{\prime \mu }=%
\frac{1}{N}\ell ^{\mu }$ \\ 
\\ 
$n_{\mu }^{\prime }=Nn_{\mu }\quad ,\quad n^{\prime \mu }=\frac{1}{\Lambda }%
n^{\mu }$ \\ 
\\ 
$m_{\mu }^{\prime }=Mm_{\mu }\quad ,\quad m^{\prime \mu }=\frac{1}{\overline{%
M}}m^{\mu }$ \\ 
\end{tabular}
\label{e1}
\end{equation}%
Under these transformations the Newman-Penrose spin coefficients\cite{CHAND}
transform as follows%
\begin{equation}
\begin{tabular}{l}
$\alpha ^{\prime }=\frac{1}{M}\alpha +\frac{M\ \overline{M}-\Lambda N}{%
4M\Lambda N}(\overline{\tau }+\pi )+\frac{1}{4M}\overline{\delta }\ln \frac{%
\Lambda }{N\overline{M}^{2}}$ \\ 
$\beta ^{\prime }=\frac{1}{\overline{M}}\beta +\frac{M\ \overline{M}-\Lambda
N}{4\overline{M}\Lambda N}(\tau +\overline{\pi })+\frac{1}{4\overline{M}}%
\delta \ln \frac{\Lambda M^{2}}{N}$ \\ 
$\gamma ^{\prime }=\frac{1}{\Lambda }\gamma +\frac{M\ \overline{M}-\Lambda N%
}{4M\ \overline{M}\Lambda }(\overline{\mu }-\mu )+\frac{1}{4\Lambda }\Delta
\ln \frac{M}{N^{2}\overline{M}}$ \\ 
$\varepsilon ^{\prime }=\frac{1}{N}\varepsilon +\frac{M\ \overline{M}%
-\Lambda N}{4M\ \overline{M}N}(\overline{\rho }-\rho )+\frac{1}{4N}D\ln 
\frac{M\Lambda ^{2}}{\overline{M}}$ \\ 
$\mu ^{\prime }=\frac{1}{2\Lambda }(\mu +\overline{\mu })+\frac{N}{2M\ 
\overline{M}}(\mu -\overline{\mu })+\frac{1}{2\Lambda }\Delta \ln (M\ 
\overline{M})$ \\ 
$\rho ^{\prime }=\frac{1}{2N}(\rho +\overline{\rho })+\frac{\Lambda }{2M\ 
\overline{M}}(\rho -\overline{\rho })-\frac{1}{2N}D\ln (M\ \overline{M})$ \\ 
$\pi ^{\prime }=\frac{\overline{M}}{2\Lambda N}(\pi +\overline{\tau })+\frac{%
1}{2M}(\pi -\overline{\tau })+\frac{1}{2M}\overline{\delta }\ln (\Lambda N)$
\\ 
$\tau ^{\prime }=\frac{M}{2\Lambda N}(\tau +\overline{\pi })+\frac{1}{2%
\overline{M}}(\tau -\overline{\pi })-\frac{1}{2\overline{M}}\delta \ln
(\Lambda N)$ \\ 
$\kappa ^{\prime }=\frac{\Lambda }{N\overline{M}}\kappa \quad ,\quad \sigma
^{\prime }=\frac{M}{N\overline{M}}\sigma \quad ,\quad \nu ^{\prime }=\frac{N%
}{\Lambda M}\nu \quad ,\quad \lambda ^{\prime }=\frac{\overline{M}}{\Lambda M%
}\lambda $%
\end{tabular}
\label{e2}
\end{equation}%
We see that the following relations 
\begin{equation}
\begin{tabular}{l}
$\rho ^{\prime }-\overline{\rho ^{\prime }}=\frac{\Lambda }{M\overline{M}}%
(\rho -\overline{\rho })$ \\ 
$\mu ^{\prime }-\overline{\mu ^{\prime }}=\frac{N}{M\overline{M}}(\mu -%
\overline{\mu })$ \\ 
$\tau ^{\prime }+\overline{\pi ^{\prime }}=\frac{M}{\Lambda N}(\tau +%
\overline{\pi })$%
\end{tabular}
\label{e3}
\end{equation}%
establish the corresponding quantities as relative invariants of the complex
structure, analogous to the Levi forms of the CR structure\cite{JACO1990}.

Not all the metrics admit a lorentzian complex structure. In that case, the
(non-conformally flat) metric uniquely determines the lorentzian complex
structure through the integrability conditions

\begin{equation}
\Psi _{ABCD}\ o^{A}o^{B}o^{C}o^{D}=0=\Psi _{ABCD}\ \iota ^{A}\iota ^{B}\iota
^{C}\iota ^{D}  \label{e4}
\end{equation}%
where $o^{A}$ and $\iota ^{A}$ is the spinor dyad of the integrable null
tetrad and $\Psi _{ABCD}$\ is the Weyl tensor in the spinor notation.\
Namely, they are principal null directions of the Weyl spinor $\Psi _{ABCD}$%
. But the inverse is not true. The class $J[g_{\mu \nu }]$ of metrics, which
are compatible with a lorentzian complex structure,\ is determined by the
following general form 
\begin{equation}
\begin{tabular}{l}
$g_{\mu \nu }^{\prime }=\phi ^{2}g_{\mu \nu }+\psi ^{2}(\ell _{\mu }n_{\nu
}+n_{\mu }\ell _{\nu })$ \\ 
\\ 
$g^{\prime \mu \nu }=\frac{1}{\phi ^{2}}g^{\mu \nu }-\frac{\psi ^{2}}{\phi
^{2}+\psi ^{2}}(\ell ^{\mu }n^{\nu }+n^{\mu }\ell ^{\nu })$%
\end{tabular}
\label{e5}
\end{equation}

This is a generalization of the conformal (Weyl) class of metrics and a
Yamabe-like problem may be posed with one more scalar quantity constant,
besides the scalar curvature. I do not actually see any physical relevance
of this mathematical problem. Instead I have already pointed out\cite%
{RAG1999},\cite{RAG2008b} that the derivation of Einstein's gravity implies
that the energy of the model should be defined using the Einstein tensor of
a metric from the class $J[g_{\mu \nu }]$.\ The assumed in General
Relativity dominant energy condition must be proven in the present model. I
want to point out that the existence of a positive conserved quantity seems
to be essential for the quantum stability of the model.

Einstein used the Levi-Civita connection to equate the covariantly conserved
tensor $E^{\mu \nu }$ with the matter energy-momentum tensor

\begin{equation}
E^{\mu \nu }\equiv R^{\mu \nu }-\frac{1}{2}\,R\,g^{\mu \nu }=8\pi k\,T^{\mu
\nu }  \label{e6}
\end{equation}%
where $T^{\mu \nu }$ is the energy-momentum tensor of the matter fields. In
the present model, general covariance does not permit the definition of an
energy-momentum tensor. Therefore we use the existence of a coordinate
system such that 
\begin{equation}
\partial _{\mu }\left( \sqrt{-g}E^{\mu \nu }\right) =0  \label{e7}
\end{equation}%
to define the conserved quantity 
\begin{equation}
\mathcal{E}(g_{\mu \nu })=\int_{\!t}\sqrt{-g}E^{\mu 0}dS_{\mu }  \label{e8}
\end{equation}%
in the precise coordinate system. Recall that singularities are not
permitted in Quantum Field Theory. Therefore we have to consider only
regular spacetime metrics.

The considered quantity $\mathcal{E}(g_{\mu \nu })$ depends on the metric $%
g_{\mu \nu }$ and it does not characterize the complex structure, therefore
it cannot be the energy definition of the lorentzian complex structure. I
think that the energy of a complex structure is properly defined by the
following minimum 
\begin{equation}
\mathrm{E}[J_{\mu }^{\;\nu }]=\underset{g_{\mu \nu }\in J[g_{\mu \nu }]}{%
\min }\mathcal{E}(g_{\mu \nu })  \label{e9}
\end{equation}%
where the minimum is taken over all the class $J[g_{\mu \nu }]$ of metrics.
Apparently the mathematical conjecture is that such a minimum exists, which
is not at all evident! That is, in the present model, the positive energy
condition must be proved!

This conserved quantity depends only on the moduli parameters of the complex
structure. Minkowski spacetime determines the vacuum sector of the model
because $\mathrm{E}[J_{\mu }^{\;\nu }]=0$, for complex structures compatible
with the Minkowski metric. From the 2-dimensional solitonic models\cite%
{FELS1981}, we know that the minima of the energy characterize the solitons.
Assuming that $\mathrm{E}[J_{\mu }^{\;\nu }]$ is a smooth function of the
moduli parameters, we can always expand it around a minimum. 
\begin{equation}
\mathrm{E}[J_{\mu }^{\;\nu }]\simeq E+\underset{q}{\sum }\varepsilon _{q}\ 
\overline{a_{q}}\ a_{q}  \label{e10}
\end{equation}%
where $E$ and $\varepsilon _{q}$ are positive parameters. These variables
and $a_{q}$ are moduli parameters of the complex structure. $E$ is defined
to be the energy of the soliton characterized by the minimum and $%
\varepsilon _{q}$ are the energies of the excitation modes.

This formal procedure implies the Einstein equations if the following points
are mathematically clarified: 1) The positivity of $\mathcal{E}(g_{\mu \nu
}) $ for at least a subclass of $J[g_{\mu \nu }]$. 2) The precise coordinate
system satisfying (\ref{e7}) properly transforms under the Poincar\'{e}
group found in my previous work\cite{RAG2008b} and which will be outlined in
the next section. 3) The proof will explicitly single out the precise
Einstein metric from all the other induced\cite{RAG2008b} metrics on the
manifold.

\section{THE\ G$_{2,2}$ DESCRIPTION OF THE LORETZIAN COMPLEX STRUCTURES}

\setcounter{equation}{0}

It is trivial to show from (\ref{i4}) that the structure coordinates $%
(z^{\alpha },\;z^{\widetilde{\alpha }})$,\ $\alpha =0,\ 1$ satisfy the
relations

\begin{equation}
\begin{array}{l}
dz^{0}\wedge dz^{1}\wedge d\overline{z^{0}}\wedge d\overline{z^{1}}=0 \\ 
\\ 
dz^{\widetilde{0}}\wedge dz^{\widetilde{0}}\wedge d\overline{z^{0}}\wedge d%
\overline{z^{1}}=0 \\ 
\\ 
dz^{\widetilde{0}}\wedge dz^{\widetilde{0}}\wedge d\overline{z^{\widetilde{0}%
}}\wedge d\overline{z^{\widetilde{0}}}=0%
\end{array}
\label{g1}
\end{equation}%
that is, there are two real functions $\Psi _{11}$ , $\Psi _{22}$ and a
complex one $\Psi _{12}$, such

\begin{equation}
\Psi _{11}(\overline{z^{\alpha }},z^{\alpha })=0\quad ,\quad \Psi
_{12}\left( \overline{z^{\alpha }},z^{\widetilde{\alpha }}\right) =0\quad
,\quad \Psi _{22}\left( \overline{z^{\widetilde{\alpha }}},z^{\widetilde{%
\alpha }}\right) =0  \label{g2}
\end{equation}%
This surface may be considered as the characteristic boundary of a domain
which is holomorphically equivalent to a bounded domain in $%
\mathbb{C}
^{4}$, through the positive definite condition of the following $2\times 2$\
matrix%
\begin{equation}
\Psi =%
\begin{pmatrix}
\Psi _{11} & \Psi _{12} \\ 
\overline{\Psi _{12}} & \Psi _{22}%
\end{pmatrix}%
>0  \label{g3}
\end{equation}%
which occurs when $\Psi _{11}+\Psi _{22}>0$ and $\det \Psi >0$.\ Notice that
the boundary conditions $\Psi _{11}+\Psi _{22}=0$ and $\det \Psi =0$ imply
the above (\ref{g2}) four relations which determine the lorentzian complex
structure.

The mathematical study of this kind of problems is performed after their
projective formulation. For this purpose I consider the rank-2 $4\times 2$
matrices $X^{mi}$\ with every column being a point of an algebraic surface $%
K_{i}(X^{mi})$ of the $CP^{3}$\ projective space. Then I consider that the $%
2\times 2$ matrix $\Psi $ has the form%
\begin{equation}
\Psi =X^{\dagger }EX-%
\begin{pmatrix}
G_{11} & G_{12} \\ 
\overline{G_{12}} & G_{22}%
\end{pmatrix}%
>0  \label{g4}
\end{equation}%
where $E$\ is an $SU(2,2)$ invariant $4\times 4$ matrix and $G_{ij}=G_{ij}(%
\overline{X^{mi}},X^{mj})$ are homogeneous functions. This projective form
emerged from the consideration\cite{RAG2008b} of lorentzian complex
structures asymptotically compatible with the Minkowski metric and the
Penrose observation\cite{P-R1984} that a geodetic and shear free congruence
of Minkowski spacetime can be described by a null twistor satisfying an
algebraic condition. In the simple case $G_{ij}=0$ it is a first kind Siegel
domain\cite{PYAT},\cite{XU} for \ 
\begin{equation}
E=%
\begin{pmatrix}
0 & I \\ 
I & 0%
\end{pmatrix}
\label{g5}
\end{equation}%
which is holomorphic to the $SU(2,2)$ invariant bounded classical domain
given by \ 
\begin{equation}
E=%
\begin{pmatrix}
I & 0 \\ 
0 & -I%
\end{pmatrix}
\label{g6}
\end{equation}%
That is the form (\ref{g4}) osculates the surface (\ref{g2}) with the Shilov
boundary of the $SU(2,2)$ invariant classical domain.

Using the following spinorial form of the rank-2 matrix $X^{mj}$ in its
ubounded realization 
\begin{equation}
\begin{array}{l}
X^{mj}=%
\begin{pmatrix}
\lambda ^{Aj} \\ 
-ir_{A^{\prime }B}\lambda ^{Bj}%
\end{pmatrix}
\\ 
\end{array}
\label{g7}
\end{equation}%
and the null tetrad 
\begin{equation}
\begin{array}{l}
L^{a}=\frac{1}{\sqrt{2}}\overline{\lambda }^{A^{\prime }1}\lambda
^{B1}\sigma _{A^{\prime }B}^{a}\quad ,\quad N^{a}=\frac{1}{\sqrt{2}}%
\overline{\lambda }^{A^{\prime }2}\lambda ^{B2}\sigma _{A^{\prime
}B}^{a}\quad ,\quad M^{a}=\frac{1}{\sqrt{2}}\overline{\lambda }^{A^{\prime
}2}\lambda ^{B1}\sigma _{A^{\prime }B}^{a} \\ 
\\ 
\epsilon _{AB}\lambda ^{A1}\lambda ^{B2}=1 \\ 
\end{array}
\label{g8}
\end{equation}%
the above relations take the form 
\begin{equation}
\begin{array}{l}
\Psi _{11}=2\sqrt{2}y^{a}L_{a}-G_{11}(\overline{Y^{m1}},Y^{n1}) \\ 
\\ 
\Psi _{12}=2\sqrt{2}y^{a}\overline{M}_{a}-G_{12}(\overline{Y^{m1}},Y^{n2})
\\ 
\\ 
\Psi _{22}=2\sqrt{2}y^{a}N_{a}-G_{22}(\overline{Y^{m2}},Y^{n2})%
\end{array}
\label{g9}
\end{equation}%
where $y^{a}$\ is the imaginary part of $r^{a}=x^{a}+iy^{a}$\ defined by the
relation $r_{A^{\prime }B}=r^{a}\sigma _{aA^{\prime }B}$\ and $\sigma
_{A^{\prime }B}^{a}$ being the identity and the three Pauli matrices. The
surface satisfies the relation \ 
\begin{equation}
\begin{array}{l}
y^{a}=\frac{1}{2\sqrt{2}}[G_{22}N^{a}+G_{11}L^{a}-G_{12}M^{a}-\overline{%
G_{12}}\overline{M}^{a}] \\ 
\end{array}
\label{g10}
\end{equation}%
which combined with the computation of $\lambda ^{Ai}$\ as functions of $%
r^{a}$, using the Kerr conditions $K_{i}(X^{mi})$, permit us to compute $%
y^{a}=$ $y^{a}(x)$ as functions of the real part of $r^{a}$.

Notice that this surface does not generally belong into the Seigel domain,
because $y^{0}$\ and \ 
\begin{equation}
\begin{array}{l}
y^{a}y^{b}\eta _{ab}=\frac{1}{8}[G_{22}G_{11}-G_{12}\overline{G_{12}}] \\ 
\end{array}
\label{g11}
\end{equation}%
are not always positive. But the regular surfaces (with an upper bound) can
always be brought inside the Siegel domain (and its holomorphic bounded
classical domain) with an holomorphic complex time translation. 
\begin{equation}
\begin{array}{l}
\begin{pmatrix}
\lambda ^{\prime Aj} \\ 
w_{B^{\prime }}^{\prime j}%
\end{pmatrix}%
=%
\begin{pmatrix}
I & 0 \\ 
dI & I%
\end{pmatrix}%
\begin{pmatrix}
\lambda ^{Aj} \\ 
w_{B^{\prime }}^{j}%
\end{pmatrix}
\\ 
\end{array}
\label{g12}
\end{equation}%
In the case of an asymptotically flat space time, its point at infinity,
which is on the Shilov boundary, remains intact. Therefore we can always
assume that the bounded domain (\ref{g4}) is always inside the $SU(2,2)$
invariant classical domain.

A typical example is the case $X^{\dagger }\Gamma X=0$ with \ 
\begin{equation}
\Gamma =%
\begin{pmatrix}
-2b & I \\ 
I & 0%
\end{pmatrix}
\label{g13}
\end{equation}%
and the Kerr functions $K_{1}=X^{11}+$\ $X^{31}=0$ and $K_{2}=X^{02}+$\ $%
X^{22}$\bigskip $=0$. Then using the following definition of the structure
coordinates 
\begin{equation}
z^{0}=i\frac{X^{21}}{X^{01}}\quad ,\quad z^{1}=\frac{X^{11}}{X^{01}}\quad
,\quad z^{\widetilde{0}}=i\frac{X^{32}}{X^{12}}\quad ,\quad z^{\widetilde{1}%
}=-\frac{X^{02}}{X^{12}}  \label{g14}
\end{equation}%
we easily find the relations 
\begin{equation}
\begin{array}{l}
\Psi _{11}=i(\overline{z^{0}}-z^{0})-2z^{1}\overline{z^{1}}-2b(1+z^{1}%
\overline{z^{1}}) \\ 
\\ 
\Psi _{12}=z^{\widetilde{1}}(1-i\overline{z^{0}}+2b)-\overline{z^{1}}(1+iz^{%
\widetilde{0}}+2b) \\ 
\\ 
\Psi _{22}=i(\overline{z^{\widetilde{0}}}-z^{\widetilde{0}})-2z^{\widetilde{1%
}}\overline{z^{\widetilde{1}}}-2b(1+z^{\widetilde{1}}\overline{z^{\widetilde{%
1}}})%
\end{array}
\label{g15}
\end{equation}%
which will be described below. We find $y^{a}=(b,0,0,0)$ which does not
induce any gravity, because this lorentzian complex structure is compatible
with the Minkowski metric.\ 

The asymptotically flat lorentzian complex structures ($X^{\dagger
1}EX^{1}=0\ ,\ X^{\dagger 2}EX^{2}=0$) belong into irreducible
representations of the $SU(2,2)$ group which is broken down exactly\cite%
{RAG2008b} to its $Poincar\acute{e}\times dilation$ subgroup by the infinity
point on the Shilov boundary.

The real part of $r^{a}$\ determine a characteristic set of coordinates
because it properly transforms under the Poincar\'{e} group. In the case of
asymptotically flat lorentzian complex structures there are two other
coordinate systems, which properly transform under the Poincar\'{e} group
too. These are the geodetic coordinates of $\ell ^{\mu }$ and $n^{\mu }$
which have the general forms 
\begin{equation}
\begin{array}{l}
x_{(+)A^{\prime }A}=\frac{iw_{A^{\prime }}^{1}\overline{w^{1}}_{A}}{\lambda
^{C1}\overline{w^{1}}_{C}}+r\overline{\lambda ^{1}}_{A^{\prime }}\lambda
_{A}^{1}\quad ,\quad \forall \ r\quad and\ \quad \lambda ^{C1}\overline{w^{1}%
}_{C}\neq 0 \\ 
\\ 
x_{(-)A^{\prime }A}=\frac{iw_{A^{\prime }}^{2}\overline{w^{2}}_{A}}{\lambda
^{C2}\overline{w^{2}}_{C}}+s\overline{\lambda ^{2}}_{A^{\prime }}\lambda
_{A}^{2}\quad ,\quad \forall \ s\quad and\ \quad \lambda ^{C2}\overline{w^{2}%
}_{C}\neq 0%
\end{array}
\label{g16}
\end{equation}%
These characteristic coordinate systems should be related to the definition
of the energy of a lorentzian complex structure, for this quantity to be a
component of a four-momentum.

\section{CLASSIFICATION OF LORENTZIAN COMPLEX STRUCTURES}

\setcounter{equation}{0}

Flaherty worked\cite{FLAHE1976} with the complex structure preserving
connection $\gamma _{bc}^{a}$ with the following non vanishing components 
\begin{equation}
\begin{tabular}{l}
$\gamma _{\beta \gamma }^{\alpha }=g^{\alpha \widetilde{\alpha }}\ \partial
_{\beta }g_{\gamma \widetilde{\alpha }}\quad ,\quad \gamma _{\widetilde{%
\beta }\widetilde{\gamma }}^{\widetilde{\alpha }}=g^{\alpha \widetilde{%
\alpha }}\ \partial _{\widetilde{\beta }}g_{\alpha \widetilde{\gamma }}$ \\ 
\end{tabular}
\label{c1}
\end{equation}%
where the metric is written in the structure coordinate system $(z^{\alpha
},\;z^{\widetilde{\alpha }})$,\ $\alpha =0,\ 1$ 
\begin{equation}
\begin{tabular}{l}
$ds^{2}=2g_{\alpha \widetilde{\beta }}\ dz^{\alpha }dz^{\widetilde{\beta }}$
\\ 
\end{tabular}
\label{c2}
\end{equation}%
He showed that if the torsion of this connection $T_{ab}^{c}=\gamma
_{ba}^{c}-\gamma _{ab}^{c}$ vanishes, the complex structure is kaehlerian, $%
d(J_{\mu \nu }\ dx^{\mu }\wedge dx^{\nu })$, and the vectors of the null
tetrad are hypersurface orthogonal. This means that the complex structure is
trivial and apparently compatible with the Minkowski metric. But the inverse
is not valid. There are non-trivial complex structures (with non-vanishing
torsion) which are also compatible with the Minkowski spacetime. Hence we
cannot use this torsion to describe the gravitational content of the complex
structure. But we may use all the invariant tensors (torsion, curvature and
their covariant derivatives) created by this connection to classify the
lorentzian complex structure.

The four real conditions (\ref{g2}) imply that the spacetime, which admits
an integrable lorentzian complex structure, is a CR manifold with
codimension four. Following the ordinary procedure we can find the
corresponding four real forms. It is convenient to use the notation $%
\partial f=\frac{\partial f}{\partial z^{\alpha }}dz^{\alpha }$\ and $%
\widetilde{\partial }f=\frac{\partial f}{\partial z^{\widetilde{\alpha }}}%
dz^{\widetilde{\alpha }}$. Then we find \ 
\begin{equation}
\begin{array}{l}
\ell =2i\partial \Psi _{11}=i(\partial -\overline{\partial })\Psi _{11}=-2i%
\overline{\partial }\Psi _{11} \\ 
\\ 
n=2i\widetilde{\partial }\Psi _{22}=i(\widetilde{\partial }-\overline{%
\widetilde{\partial }})\Psi _{22}=-2i\overline{\widetilde{\partial }}\Psi
_{22} \\ 
\\ 
m_{1}=i(\partial +\widetilde{\partial }-\overline{\partial }-\overline{%
\widetilde{\partial }})\frac{\Psi _{12}+\overline{\Psi _{12}}}{2} \\ 
\\ 
m_{2}=i(\partial +\widetilde{\partial }-\overline{\partial }-\overline{%
\widetilde{\partial }})\frac{\overline{\Psi _{12}}-\Psi _{12}}{2i} \\ 
\end{array}
\label{c3}
\end{equation}

These forms restricted on the manifold are real, because of $d\Psi _{ij}=0$
and the special dependence of each function on the structure coordinates $%
\left( z^{\alpha },z^{\widetilde{\alpha }}\right) $. The relations become
simpler if we use the complex form \ 
\begin{equation}
\begin{array}{l}
m=m_{1}+im_{2}=2i\partial \overline{\Psi _{12}}=-2i\overline{\widetilde{%
\partial }}\overline{\Psi _{12}}=i(\partial -\overline{\widetilde{\partial }}%
)\overline{\Psi _{12}} \\ 
\end{array}
\label{c4}
\end{equation}%
Notice that these forms coincide with the null tetrad up to a multiplicative
factor. The general CR transformation is actually restricted to a factor,
because the dimension of the manifold coincides with its codimension.

A general null tetrad has the following differential forms \ 
\begin{equation}
\begin{array}{l}
d\ell =(\varepsilon +\overline{\varepsilon })n\wedge \ell +(\overline{\tau }%
-\alpha -\overline{\beta })m\wedge \ell +(\tau -\overline{\alpha }-\beta )%
\overline{m}\wedge \ell + \\ 
\qquad +(\rho -\overline{\rho })m\wedge \overline{m}-\overline{\kappa }%
n\wedge m-\kappa n\wedge \overline{m} \\ 
\\ 
dn=-(\gamma +\overline{\gamma })\ell \wedge n+(\alpha +\overline{\beta }-\pi
)m\wedge n+(\overline{\alpha }+\beta -\overline{\pi })\overline{m}\wedge n+
\\ 
\qquad +(\mu -\overline{\mu })m\wedge \overline{m}+\nu \ell \wedge m+%
\overline{\nu }\ell \wedge \overline{m} \\ 
\\ 
dm=(\gamma -\overline{\gamma }+\overline{\mu })\ell \wedge m+(\varepsilon -%
\overline{\varepsilon }-\rho )n\wedge m+(\overline{\alpha }-\beta )\overline{%
m}\wedge m- \\ 
\qquad -(\tau +\overline{\pi })\ell \wedge n+\overline{\lambda }\ell \wedge 
\overline{m}-\sigma n\wedge \overline{m} \\ 
\end{array}
\label{c5}
\end{equation}%
It is integrable if $\kappa =\sigma =\lambda =\nu =0$.\ Then the
transformations of the spin coefficients (\ref{e2}) imply that the vanishing
or not of the quantities $(\rho -\overline{\rho })\ ,(\mu -\overline{\mu })\
,\ (\tau +\overline{\pi })$ are relative invariants of the lorentzian
complex structure. If these quantities vanish, the complex structure is
kaehlerian, and the vectors of the null tetrad are hypersurface orthogonal.
That is the complex structure is trivial and apparently compatible with the
Minkowski metric.

The classification of the lorentzian complex structures may be approached
using the CR structure techniques. In the next two subsections I will
outline the Chern-Moser normal form and the $SU(2,2)$ Cartan connection
methods.

\subsection{The Chern-Moser normal form method}

It has already pointed out\cite{N-N2005},\cite{A-N2009} that the explicit
conditions $\Psi _{11}(\overline{z^{\alpha }},z^{\alpha })=0\ ,\ \Psi
_{22}\left( \overline{z^{\widetilde{\alpha }}},z^{\widetilde{\alpha }%
}\right) =0$\ and the corresponding holomorphic transformations $z^{\prime
\alpha }=f^{\alpha }(z^{\alpha })$ and $z^{\prime \widetilde{\alpha }}=f^{%
\widetilde{\alpha }}(z^{\widetilde{\alpha }})$ which preserve the lorentzian
complex structure, are exactly those of the 3-dimensional CR structures\cite%
{JACO1990}. Therefore we may use the Moser procedure for the classification
of the lorentzian complex structures. For each hypersurface type CR
structure we consider the following Moser expansions \ 
\begin{equation}
\begin{array}{l}
U=z^{1}\overline{z^{1}}+\tsum\limits_{k\geq 2,j\geq 2}N_{jk}(u)(z^{1})^{j}(%
\overline{z^{1}})^{k} \\ 
N_{22}=N_{32}=N_{33}=0 \\ 
V=z^{\widetilde{1}}\overline{z^{\widetilde{1}}}+\tsum\limits_{k\geq 2,j\geq
2}\widetilde{N}_{jk}(v)(z^{\widetilde{1}})^{j}(\overline{z^{\widetilde{1}}}%
)^{k} \\ 
\widetilde{N}_{22}=\widetilde{N}_{32}=\widetilde{N}_{33}=0%
\end{array}
\label{c6}
\end{equation}%
where $z^{0}=u+iU\ ,\ z^{\widetilde{0}}=v+iV$\ and the functions $N_{jk}(u)\
,\ \widetilde{N}_{jk}(v)$ characterize the lorentzian complex structure. By
their construction these functions belong into representations of the
isotropy subgroup of $SU(1,2)$ symmetry group of the hyperquadric. Notice
that the corresponding Moser chains are determined by $n^{\alpha }\frac{%
\partial }{\partial z^{\alpha }}$ and $\ell ^{\widetilde{\alpha }}\frac{%
\partial }{\partial z\widetilde{^{\alpha }}}$\ respectively and they should
be related to $x_{\pm }^{a}$\ geodetic coordinates.

The above Moser normal forms are unique up to the isotropy group of the
hyperquadric for each expansion. The transformations of the isotropy
subgroup have the form \ 
\begin{equation}
\begin{array}{l}
z^{\prime 1}=\frac{\overline{c}(z^{1}+az^{0})}{c^{2}[1-2i\overline{a}%
z^{1}+(b-i|a|^{2})z^{0}]} \\ 
\\ 
z^{\prime 0}=\frac{z^{0}}{c\overline{c}[1-2i\overline{a}%
z^{1}+(b-i|a|^{2})z^{0}]} \\ 
\end{array}
\label{c7}
\end{equation}%
where the parameters $a,c$ are complex and $b$ is real. An analogous
transformation ambiguity exists for the tilded structure coordinates. This
freedom may be used to fix the linear terms of the $\Psi _{12}=0$ expansion
as follows \ 
\begin{equation}
\begin{array}{l}
z^{\widetilde{1}}=\overline{z^{1}}+C_{1}(z^{\widetilde{0}})^{2}+C_{2}(%
\overline{z^{0}})^{2}+C_{3}(\overline{z^{1}})^{2}+.... \\ 
\end{array}
\label{c8}
\end{equation}

The Moser normal forms, which determine a complex structure compatible with
the Minkowski metric (without gravity content), may be found using the
condition $X^{\dagger }EX=0$\ and the two Kerr algebraic homogeneous
conditions $K_{i}(X^{mi})=0$.

\subsection{The Cartan connection method}

The osculation (\ref{g4}) of the of the CR structure which describes the
integrable lorentzian complex structure suggests the use of the $U(2,2)$\
Cartan connection. The Cartan connection of a group manifold is $\omega
=g^{-1}dg$,\ where $g$\ is a $4\times 4$\ matrix, which preserves the form
of $E$, ($g^{\dagger }Eg=E$)\ and it has the following coset space
decomposition \ 
\begin{equation}
\begin{array}{l}
g=%
\begin{pmatrix}
I & 0 \\ 
-ix & I%
\end{pmatrix}%
\begin{pmatrix}
\lambda & i\lambda b \\ 
0 & (\lambda ^{\dagger })^{-1}%
\end{pmatrix}
\\ 
\end{array}
\label{c15}
\end{equation}%
where $\lambda $ is a general complex and $x,b$ are hermitian $2\times 2$
matrices. The curvature of this connection is $\Omega =d\omega +\omega
\wedge \omega =0$.\ It is known that not all these $g$\ matrices determine a
complex structure of the Minkowski spacetime. They do, if the matrix $%
\lambda $ has the form \ 
\begin{equation}
\begin{array}{l}
\lambda =%
\begin{pmatrix}
k & -\widetilde{w}\widetilde{k} \\ 
wk & \widetilde{k}%
\end{pmatrix}
\\ 
\end{array}
\label{c16}
\end{equation}%
where $k,\widetilde{k}$ are complex and $w,\widetilde{w}$ are functions of $%
x_{A^{\prime }A}$\ implied by the Kerr conditions $K_{1}(w,x_{0^{\prime
}0}+x_{0^{\prime }1}w,x_{1^{\prime }0}+x_{1^{\prime }1}w)=0$ and $K_{2}(%
\widetilde{w},-x_{0^{\prime }0}\widetilde{w}+x_{0^{\prime }1},-x_{1^{\prime
}0}\widetilde{w}+x_{1^{\prime }1})=0$. The curvature of this lift of
Minkowski spacetime continues to vanish, because $g$ is still an element of $%
U(2,2)$.

In the general case (\ref{g4}) of a lorentzian complex structure which is
not compatible with the Minkowski metric, the hermitian matrix $x_{A^{\prime
}A}$ is replaced by the general complex matrix $r_{A^{\prime }A}(x)$,\ which
is determined by the surface defining conditions. Then $g$ is no longer an
element of $U(2,2)$\ and the corresponding curvature does not vanish.\ 

In the general case the Cartan connection is \ 
\begin{equation}
\begin{array}{l}
\omega =%
\begin{pmatrix}
e_{1} & ie_{2} \\ 
-ie_{0} & -e_{1}^{\dagger }%
\end{pmatrix}
\\ 
\end{array}
\label{c17}
\end{equation}%
where $e_{0}$\ and $e_{2}$ are $2\times 2$ hermitian matrices 1-forms.
Identifying \ 
\begin{equation}
\begin{array}{l}
e_{0}=%
\begin{pmatrix}
\ell & \overline{m} \\ 
m & n%
\end{pmatrix}
\\ 
\end{array}
\label{c18}
\end{equation}%
the curvature $\Omega $ satisfies the relations \ 
\begin{equation}
\begin{array}{l}
de_{0}+e_{0}\wedge e_{1}-e_{1}^{\dagger }\wedge e_{0}=i\Omega _{12} \\ 
\\ 
de_{1}+e_{1}\wedge e_{1}+e_{2}\wedge e_{0}=\Omega _{11} \\ 
\\ 
de_{2}+e_{1}\wedge e_{2}-e_{2}\wedge e_{1}^{\dagger }=-i\Omega _{21} \\ 
\end{array}
\label{c19}
\end{equation}%
with \ 
\begin{equation}
\begin{array}{l}
e_{1}=%
\begin{pmatrix}
\gamma & \nu \\ 
-\tau & -\gamma%
\end{pmatrix}%
\ell +%
\begin{pmatrix}
\varepsilon & \pi \\ 
-\kappa & -\varepsilon%
\end{pmatrix}%
n+ \\ 
\qquad +%
\begin{pmatrix}
-\alpha & -\lambda \\ 
\rho & \alpha%
\end{pmatrix}%
m+%
\begin{pmatrix}
-\beta & -\mu \\ 
\sigma & \beta%
\end{pmatrix}%
\overline{m} \\ 
\\ 
e_{2}=%
\begin{pmatrix}
-\Phi _{22} & \Phi _{21} \\ 
\Phi _{12} & \Lambda -\Phi _{11}%
\end{pmatrix}%
\ell +%
\begin{pmatrix}
\Lambda -\Phi _{11} & \Phi _{10} \\ 
\Phi _{01} & -\Phi _{00}%
\end{pmatrix}%
n+ \\ 
\qquad +%
\begin{pmatrix}
\Phi _{21} & -\Phi _{20} \\ 
-\Lambda -\Phi _{11} & \Phi _{10}%
\end{pmatrix}%
m+%
\begin{pmatrix}
\Phi _{12} & -\Lambda -\Phi _{11} \\ 
-\Phi _{02} & \Phi _{01}%
\end{pmatrix}%
\overline{m} \\ 
\end{array}
\label{c20}
\end{equation}%
and the curvature components are $\Omega _{12}=0$,\ \ 
\begin{equation}
\begin{array}{l}
\Omega _{11}=%
\begin{pmatrix}
-\Psi _{2} & -\Psi _{3} \\ 
\Psi _{1} & \Psi _{2}%
\end{pmatrix}%
\ell \wedge n+%
\begin{pmatrix}
\Psi _{3} & \Psi _{4} \\ 
-\Psi _{2} & -\Psi 3%
\end{pmatrix}%
n\wedge m+ \\ 
\qquad +%
\begin{pmatrix}
-\Psi _{1} & -\Psi _{2} \\ 
\Psi _{0} & \Psi _{1}%
\end{pmatrix}%
n\wedge \overline{m}+%
\begin{pmatrix}
\Psi _{2} & \Psi _{3} \\ 
-\Psi _{1} & -\Psi _{2}%
\end{pmatrix}%
m\wedge \overline{m} \\ 
\end{array}
\label{c21}
\end{equation}%
and $\Omega _{21}$ is directly computed.

In the present case of integrable complex structures we have $\Psi
_{0}=0=\Psi _{4}$ and the corresponding codimension-4 CR structures are
classified to the following four cases\ \ 
\begin{equation}
\begin{array}{l}
Case\ I:\Psi _{1}\neq 0\ ,\ \Psi _{2}\neq 0\ ,\ \Psi _{3}\neq 0 \\ 
\\ 
Case\ II:\Psi _{1}\neq 0\ ,\ \Psi _{2}\neq 0\ ,\ \Psi _{3}=0 \\ 
\\ 
Case\ III:\Psi _{1}\neq 0\ ,\ \Psi _{2}=0\ ,\ \Psi _{3}=0 \\ 
\\ 
Case\ D:\Psi _{1}=0\ ,\ \Psi _{2}\neq 0\ ,\ \Psi _{3}=0 \\ 
\end{array}
\label{c22}
\end{equation}%
Notice that this classification is also related to the number of principal
null directions that the spacetime admits through the relation (\ref{e4}).

\section{A FEFFERMAN-LIKE METRIC}

\setcounter{equation}{0}

It is well known\cite{XU} that for any proper bounded domain (and its
holomorphic transformations) there is a Bergman kernel%
\begin{equation}
\begin{array}{l}
K(z,\overline{w})=\tsum\limits_{j}\phi _{j}(z)\overline{\phi _{j}(w)} \\ 
\end{array}
\label{f1}
\end{equation}%
where $[\phi _{j}(z)]$ is a complete orthonormal system of $L^{2}$\
holomorphic functions.\ For homogeneous classical domains it can be
computed. For the 4(real)-dimensional ball, the Bergman function $%
K_{H}(z)=K_{H}(z,\overline{z})$ is \ 
\begin{equation}
\begin{array}{l}
K_{H}(z)=\frac{2}{\pi ^{2}}\rho ^{-3} \\ 
\end{array}
\label{f2}
\end{equation}%
where $\rho (z,\overline{z})>0$ is the defining condition. In this case the
behavior of the Bergman function is simple as $z$\ tends to the boundary.
For the general case of a strictly pseudoconvex domain, the Bergman function
has the same leading singularity, but it has a logarithmic singularity too \ 
\begin{equation}
\begin{array}{l}
K_{H}(z)=(\phi _{0}+\phi _{1}\rho +\phi _{2}\rho ^{2})\rho ^{-3}+\psi \log
\rho \ +\phi  \\ 
\end{array}
\label{f3}
\end{equation}%
where all the functions are regular on the boundary. Fefferman\cite{FE} has
shown that these functions may be asymptotically computed using the metric \ 
\begin{equation}
\begin{array}{l}
ds_{F1}^{2}=2\tsum\limits_{j,k=0}^{2}\frac{\partial ^{2}(|z_{0}|^{2}\rho )}{%
\partial z_{j}\partial \overline{z_{k}}}dz_{j}d\overline{z_{k}} \\ 
\end{array}
\label{f4}
\end{equation}%
restricted on $S^{1}\times Boundary$, where $S^{1}$ is the natural bundle $%
|z_{0}|=1$.

The condition $\Psi _{11}(\overline{z^{\alpha }},z^{\alpha })=0$ determines
a 3-dimensional CR submanifold of the spacetime. The corresponding Fefferman
metric cannot be identified with the spacetime metric because the Fefferman
metric is always Petrov type N, while our spacetime metric admits at least
two different principal null directions. But the Fefferman metric may be
considered as an asymptotic approximation of the spacetime metric in an
appropriate coordinate system. It can also be characterized among all the
metrics of a 4-dimensional manifold using (among other restrictions)\cite%
{SPA} the positivity of the Einstein tensor component $E^{\mu \nu }K_{\mu
}K_{\nu }$, where $K^{\mu }$\ is the tangent vector of $S^{1}$.

Let us now proceed to define a Fefferman-like metric for the codimension-4
CR structure of the lorentzian complex structure. We will essentially
osculate the Shilov boundary of the $SU(2,2)$ invariant classical domain. In
this case the Bergman function is\cite{HUA} \ 
\begin{equation}
\begin{array}{l}
K_{B}(Z)=\frac{1}{V}[\det (1-Z^{\dag }Z)]^{-4} \\ 
\end{array}
\label{f5}
\end{equation}%
where the place of the defining function $\rho (z,\overline{z})$ takes the $%
\det (1-Z^{\dag }Z)$.\ Hence in complete analogy to the Fefferman procedure
I consider the following Kaehler metric \ 
\begin{equation}
\begin{array}{l}
ds_{F4}^{2}=2\tsum\limits_{J,K}\frac{\partial ^{2}(\det \Psi )}{\partial
z_{J}\partial \overline{z_{K}}}dz_{J}d\overline{z_{K}} \\ 
\end{array}
\label{f6}
\end{equation}%
where $\Psi $ is the $2\times 2$ matrix (\ref{g3})\ which defines the
spacetime as a CR manifold. A straightforward calculation gives ($f_{JK}=%
\frac{\partial ^{2}(\det \Psi )}{\partial z_{J}\partial \overline{z_{K}}}$)
\ 
\begin{equation}
\begin{array}{l}
f_{\beta \overline{\alpha }}=\Psi _{22}\frac{\partial ^{2}\Psi _{11}}{%
\partial z^{\beta }\partial \overline{z^{\alpha }}}-\frac{\partial \overline{%
\Psi _{12}}}{\partial z^{\beta }}\frac{\partial \Psi _{12}}{\partial 
\overline{z^{\alpha }}} \\ 
\\ 
f_{\beta \overline{\widetilde{\alpha }}}=\frac{\partial \Psi _{11}}{\partial
z^{\beta }}\frac{\partial \Psi _{22}}{\partial \overline{z^{\widetilde{%
\alpha }}}}-\Psi _{12}\frac{\partial ^{2}\overline{\Psi _{12}}}{\partial
z^{\beta }\partial \overline{z^{\widetilde{\alpha }}}} \\ 
\\ 
f_{\widetilde{\beta }\overline{\alpha }}=\frac{\partial \Psi _{22}}{\partial
z^{\widetilde{\beta }}}\frac{\partial \Psi _{11}}{\partial \overline{%
z^{\alpha }}}-\overline{\Psi _{12}}\frac{\partial ^{2}\Psi _{12}}{\partial
z^{\widetilde{\beta }}\partial \overline{z^{\alpha }}} \\ 
\\ 
f_{\widetilde{\beta }\overline{\widetilde{\alpha }}}=\Psi _{11}\frac{%
\partial ^{2}\Psi _{22}}{\partial z^{\widetilde{\beta }}\partial \overline{%
z^{\widetilde{\alpha }}}}-\frac{\partial \Psi _{12}}{\partial z^{\widetilde{%
\beta }}}\frac{\partial \overline{\Psi _{12}}}{\partial \overline{z^{%
\widetilde{\alpha }}}} \\ 
\end{array}
\label{f8}
\end{equation}%
which on the characteristic boundary $\Psi =0$ takes the form \ 
\begin{equation}
\begin{array}{l}
ds_{F4}^{2}|_{M}=(\ell _{\alpha }n_{\widetilde{\beta }}-m_{\alpha }\overline{%
m}_{\widetilde{\beta }})dz^{\alpha }dz^{\widetilde{\beta }} \\ 
\end{array}
\label{f9}
\end{equation}%
which is the spacetime metric written in structure coordinates. The
ambiguity factors of the null tetrads are hidden in the functions $\Psi
_{ij}=0$, which can always be multiplied with non vanishing factors, which
do not affect the characteristic boundary. Notice that the null geodesics of
this metric project on the Chern-Moser chains of the two hypersurface-type
CR submanifolds, which is a characteristic property of the Fefferman metric.
It would be interesting to see whether the present Fefferman-like metric
plays the same role to the asymptotic computation of the Bergman kernel of (%
\ref{g4}), like the ordinary Fefferman metric does in the case of
hypesurface type CR manifolds.

As a typical example, I will now compute the metric (\ref{f6}) for the flat
lorentzian complex structures generally given by the CR conditions $\Psi
_{ij}=f_{ij}\overline{X^{i}}EX^{j}=0$, where $f_{ij}$ are appropriate
factors such that the metric becomes Minkowski on the surface. For a change,
I will use the Newman complex trajectory\cite{A-N-K2009} condition to
specify the geodetic and shear free character of the congruences, which in
the present formalism takes the form\cite{RAG2008b} 
\begin{equation}
\begin{array}{l}
X^{mj}=%
\begin{pmatrix}
\lambda ^{Aj} \\ 
-ir_{A^{\prime }B}\lambda ^{Bj}%
\end{pmatrix}%
=%
\begin{pmatrix}
\lambda ^{A1} & \lambda ^{A2} \\ 
-i\xi _{A^{\prime }B}^{(1)}\lambda ^{B1} & -i\xi _{A^{\prime
}B}^{(2)}\lambda ^{Bj}%
\end{pmatrix}
\\ 
\end{array}
\label{f10}
\end{equation}%
where $\xi ^{a(i)}(\tau _{i})$\ are two generally independent trajectories.
In the case of the assumption of the Kerr-Penrose conditions $K_{i}(X^{mi})=0
$,\ the computational steps are analogous. This condition implies (ordinary)
holomorphic transformations between the structure coordinates $z^{0}=\tau
_{1}=z^{0}(r^{a})$, $z^{1}=\frac{X^{11}}{X^{01}}=z^{1}(r^{a})$, $z^{%
\widetilde{0}}=\tau _{2}=z^{\widetilde{0}}(r^{a})$, $z^{\widetilde{1}}=-%
\frac{X^{02}}{X^{12}}=z^{\widetilde{1}}(r^{a})$ and the complex variables $%
r^{a}$. The next step is to find factors $f_{ij}$ such that form\cite%
{RAG2008b} 
\begin{equation}
\begin{array}{l}
\Psi =\frac{1}{\sqrt{2}}%
\begin{pmatrix}
i(\overline{r^{a}}-r^{a})L_{a} & i(\overline{r^{a}}-r^{a})\overline{M}_{a}
\\ 
i(\overline{r^{a}}-r^{a})M_{a} & i(\overline{r^{a}}-r^{a})N_{a}%
\end{pmatrix}
\\ 
\end{array}
\label{f11}
\end{equation}%
where  (\ref{g8}) defines the flat null tetrad. Then the metric takes the
form \ 
\begin{equation}
\begin{array}{l}
ds_{F4}^{2}=\frac{1}{2}\tsum\limits_{a,b}\frac{\partial ^{2}(-(\overline{%
r^{c}}-r^{c})^{2})}{\partial r^{a}\partial \overline{r^{b}}}dr^{a}d\overline{%
r^{b}}=\eta _{ab}dr^{a}d\overline{r^{b}} \\ 
\end{array}
\label{f12}
\end{equation}%
which apparently becomes the Minkowski metric on the surface $y^{a}=\func{Im}%
(r^{a})=0$.

\section{ON THE U(2) AND\ POINCARE\ SYMMETRIES}

\setcounter{equation}{0}

In section III we showed that a regular 4-dimensional surface can always be
transferred inside the $SU(2,2)$ classical domain with an (holomorphic)
complex time translation. Therefore we may be constrained to the regular
surfaces (\ref{g4}) inside the $SU(2,2)$ classical domain and the flat
surface on its characteristic boundary. As quantum configurations, these
surfaces must belong to irreducible representation of the $SU(2,2)$ group.

The physically interesting asymptotically flat spacetimes, which admit a
lorentzian complex structure, are equivalent\cite{RAG2008b} to open surfaces
with a point (the Penrose $i^{0}$ point, where scri+ and scri- meet) at the
Shilov boundary. This point breaks $SU(2,2)$ group down to its $Poincar%
\acute{e}\times Dilation$ subgroup, which is the isotropy group of the
boundary\cite{PYAT}.\ 

In the case of the bounded realization of the $SU(2,2)$ classical domain (%
\ref{g6}) we represent the rank-2 matrix $X^{mi}$ as 
\begin{equation}
\begin{array}{l}
X=%
\begin{pmatrix}
T \\ 
zT%
\end{pmatrix}
\\ 
\end{array}
\label{p1}
\end{equation}%
where the $2\times 2$ matrices $r$ of the unbounded realization (\ref{g7})\
and $z$ are related with 
\begin{equation}
\begin{array}{l}
r=i(I+z)(I-z)^{-1} \\ 
\end{array}
\label{p2}
\end{equation}%
which implies that the point $z=I$ of the characteristic boundary of the
bounded realization of the homogeneous domain is mapped to the infinity of
the corresponding Siegel domain. In the bounded realization, a general $%
SU(2,2)$\textbf{\ }transformation is%
\begin{equation}
\begin{array}{l}
\begin{pmatrix}
T^{\prime } \\ 
z^{\prime }T^{\prime }%
\end{pmatrix}%
=%
\begin{pmatrix}
A_{11} & A_{12} \\ 
A_{21} & A_{22}%
\end{pmatrix}%
\left( 
\begin{array}{c}
T \\ 
zT%
\end{array}%
\right) \\ 
\\ 
z^{\prime }=\left( A_{21}+A_{22}\ z\right) \left( A_{11}+A_{12}\ z\right)
^{-1} \\ 
\end{array}
\label{p3}
\end{equation}%
where the $2\times 2$ matrices $A_{ij}$ must satisfy the conditions%
\begin{equation}
\begin{array}{l}
A_{11}^{\dagger }A_{11}-A_{21}^{\dagger }A_{21}=I\quad ,\quad
A_{11}^{\dagger }A_{12}-A_{21}^{\dagger }A_{22}=0\quad ,\quad
A_{22}^{\dagger }A_{22}-A_{12}^{\dagger }A_{12}=I \\ 
\end{array}
\label{p4}
\end{equation}

The $z=I$ stability subgroup $P_{I}$ must satisfy 
\begin{equation}
\begin{array}{l}
A_{21}+A_{22}=A_{11}+A_{12} \\ 
\end{array}
\label{p5}
\end{equation}%
which makes the last condition of (\ref{p4}) a simple identity. This
isotropy subgroup is in fact the bounded realization $Poincar\acute{e}\times
Dilation$ subgroup, which becomes a linear transformation in its unbounded
(Siegel domain) realization. Hence the open surfaces "hanging" from a fixed
point of the boundary belong to representations of the Poincar\'{e} group.

The characteristic boundary of the bounded $SU(2,2)$ homogeneous domain is\
the $U(2)$ manifold. Under a general $U(2,2)$ transformation the Poincar\'{e}
representation of surfaces of the $z=I$ point will be transformed to the%
\begin{equation}
\begin{array}{l}
U=\left( A_{21}+A_{22}\right) \left( A_{11}+A_{12}\right) ^{-1} \\ 
\end{array}
\label{p6}
\end{equation}%
point of $U(2)$. Two $U_{1}$ and $U_{2}$ points of the boundary are always
connected with the $U(2)$ group transformation $u$%
\begin{equation}
\begin{array}{l}
\begin{pmatrix}
T_{2} \\ 
U_{2}T_{2}%
\end{pmatrix}%
=%
\begin{pmatrix}
I & 0 \\ 
0 & U_{2}U_{1}^{-1}%
\end{pmatrix}%
\begin{pmatrix}
T_{1} \\ 
U_{1}T_{1}%
\end{pmatrix}
\\ 
\end{array}
\label{p7}
\end{equation}%
Therefore the corresponding isotropy subgroups are isomorphic, because $%
P_{2}=u\cdot P_{1}\cdot u^{-1}$. If in the present Quantum Field Theoretic
model the vacuum is the Minkowski part of the Shilov boundary\cite{RAG2008b}
with a precise infinity point $i^{0}$, the $SU(2,2)$ symmetry will be
spontaneously broken to the Poincar\'{e} subgroup. The dilation group is
also expected to be broken. The $U(2)$ group, which transfers the Poincar%
\'{e} representations of the $z=I$ infinity point to the other points of the
boundary, will be broken too.

It is well known\cite{P-R1984} that Minkowski spacetime is topologically
different than Kerr-Newman type of spacetimes, because the former can be
compactified through the continuity of its geodesics at scri, while in the
latter spacetimes it is obstructed by the mass. Current phenomenology
indicates that Poincar\'{e} representations of the broken $U(2)$ modes of
surfaces homotopic to Minkowski spacetime should be vector and scalar
(Higgs) bosonic fields. The Kerr-Newman type solitonic sector should appear
like the electronic multiplet of the Standard Model. I think that all these
points will be clarified when an energy operator compatible with the Poincar%
\'{e} subgroup of $SU(2,2)$ symmetry is found.

\end{document}